# Modernization of Professional Training of Electromechanics Bachelors: ICT-based Competence Approach

Yevhenii O. Modlo[1][0000-0003-2037-1557], Serhiy O. Semerikov[2][0000-0003-0789-0272]
and Ekaterina O. Shmeltzer[1]

[1] Kryvyi Rih Metallurgical Institute of the National Metallurgical Academy of Ukraine,
5, Stephana Tilhy St., Kryvyi Rih, 50006, Ukraine
`eugenemodlo@gmail.com`
[2] Kryvyi Rih State Pedagogical University, 54, Gagarina Ave., Kryvyi Rih, 50086, Ukraine
`semerikov@gmail.com`

**Abstract.** Analysis of the standards for the preparation of electromechanics in Ukraine showed that the electromechanic engineer is able to solve complex specialized problems and practical problems in a certain area of professional activity or in the process of study. These problems are characterized by complexity and uncertainty of conditions. The main competencies include social-personal, general-scientific, instrumental, general-professional and specialized-professional. A review of scientific publications devoted to the training of electromechanics has shown that four branches of engineering are involved in the training of electromechanical engineers: mechanical and electrical engineering (with a common core of electromechanics), electronic engineering and automation. The common use of the theory, methods and means of these industries leads to the emergence of a combined field of engineering – mechatronics. Summarizing the experience of electrical engineers professional training in Ukraine and abroad makes it possible to determine the main directions of their professional training modernization.

## 1  Introduction

The professional training of electromechanics bachelors in higher educational institutions of Ukraine is carried out in 38 universities of Ukraine within the knowledge sector 0507 – electrical engineering and electromechanics (from 1 Sept. 2015 – within knowledge sector 14 – electrical engineering). Now the licensed volume of admission to the bachelor's degree in electromechanics is 6065 students, the state order is 1108, the number of applicants enrolled in the first year is 1217. The direction of "Electromechanics" training is one of the few, according to which in 2012 the excess of the number entrants enrolled on the first year the volume of the state order (more than 10% more than the state orders volume). The related direction "Electrical engineering and electrotechnology" is also state and socially significant (Fig. 1). According to the Resolution of the Cabinet of Ministers of Ukraine No. 266 dated April 29, 2015, these directions are united in specialty 141 "Electricity, electrical engineering and electromechanics".

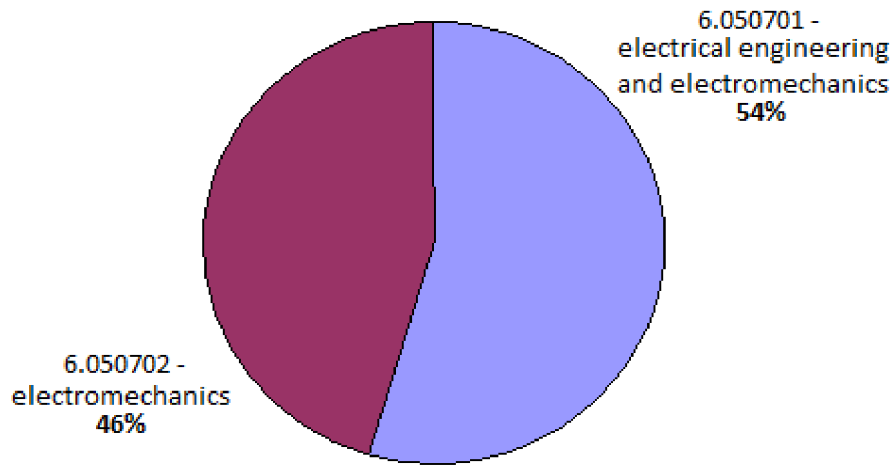

**Fig. 1.** The value of the state order in the areas of knowledge preparation 0507 – electrical engineering and electromechanics (according to [3])

## 2 General and professional competence of electrical engineers in Ukraine

The components of the sectoral standard of higher education in Ukraine (educational-professional program [22] and educational qualification characteristic [23] and ways of diagnosing the quality of higher education) are approved by the Order of the Ministry of Education and Science of Ukraine dated November 12, 2014, No. 1308. According to Educational qualification characteristics of the bachelor of electromechanics, graduates of the bachelor's degree have the qualification 2149.2 – junior electrical engineer with a generalized object of activity – "electric machines and apparatuses, electric drives, electric transport, electromechanics and systems, complexes, devices and equipment" [22, p. 6]. According to [5], electromechanical engineers have to be prepared for the development, maintenance and installation of automated, servomechanical and other electromechanical systems, in particular testing of prototype equipment, production and operational tests, system analysis, maintenance procedures, reports preparation.

Native bachelor of electromechanics should be prepared for these types of work in the processing industry field:

1. electric equipment production: electric motors, generators, transformers, distribution and control equipment, electric household appliances and other electric equipment;
2. production of machinery and general purpose equipment: engines and turbines, hydraulic and pneumatic equipment, bearings, gearing, mechanical gears and drives, lifting and handling equipment, manual electromechanical and pneumatic tools, industrial refrigeration and ventilation equipment;
3. metal-working machinery and machine tools production, machinery and metallurgical equipment, mining and construction, food and beverage manufacturing, tobacco processing, textile, sewing, fur and leather goods, paper and cardboard, plastics and rubber;
4. motor vehicles manufacturing, trailers and semitrailers: units, parts and accessories for motor vehicles, electric and electronic equipment, etc.;
5. other vehicles manufacturing: vessels and floating structures construction, pleasure and sports boats, railway locomotives and rolling stock, military vehicles, other vehicles and equipment;
6. repair and installation of machinery and equipment: repair and maintenance of finished metal products, machinery and equipment of industrial purpose, electrical equipment, ships and boats, other vehicles, other machinery and equipment, installation and installation of machinery and equipment.

According to Level 6 of the National Qualification Framework, a junior electromechanic is able to solve complex specialized problems and practical problems in a particular field of professional activity or in the process of learning that involves the application of certain theories and methods of the corresponding science and it's characterized by complexity and uncertainty of the conditions.

The description of the qualification level of the bachelor of electromechanics includes:

1. *knowledge*:

— conceptual knowledge gained in the process of learning and professional activity, including certain knowledge of contemporary achievements;
— critical understanding of the basic theories, principles, methods and concepts in teaching and professional activities;

2. *abilities*:

— solving unpredictable tasks and problems in specialized areas of professional activity or training, which involves the collection and interpretation of information (data), the choice of methods and tools, the application of innovative approaches;

3. *communication skills*:

— reporting to specialists and non-specialists of information, ideas, problems, decisions and own experience in the field of professional activity;
— the ability to effectively formulate a communication strategy;

4. *ability of autonomy and responsibility*:

- complex actions or projects manegement, responsibility and decision-making in unpredictable conditions;
- responsibility for the professional development of individuals and / or groups of people;
- ability to further study with a high level of autonomy.

For this purpose the bachelor of electromechanics should acquire the following production functions [23, p. 14-15]:

- *research* – aimed at the collection, processing, analysis and systematization of scientific and technical information on the direction of work and its use for creative decision-making of research tasks on the basis of scientific and heuristic methods);
- *design (design and development)* – the function is aimed at carrying out a purposeful sequence of actions for the synthesis of systems or their individual components, the development of documentation necessary for the implementation and use of objects and processes;
- *organizational* – is aimed at streamlining the structure and interaction of the constituent elements of the system in order to reduce uncertainty, as well as increase the efficiency of the use of resources and time;
- *managerial* – aimed at achieving the goal, ensuring the sustainable functioning and development of systems through information exchange;
- *technological* – aimed at realizing the goal of known algorithms;
- *control* – is aimed at exercising control within the scope of its professional activities in the scope of official duties;
- *prognostic* – a function that provides the opportunity, on the basis of analysis and synthesis, to carry out predictions in professional activity;
- *technical* – aimed at performing technical work in professional activities.

The junior electrician is also able to perform the following professional work:

- professionals in electric engineering field: Major Electromechanical Captain, Major Electromechanic-Commander, Power-Engineer;
- professionals in other engineering fields: electrician, junior electrician, mining engineer, engineer, engineer for the introduction of new equipment and technology, engineer for system management and maintenance, engineer-designer, repair engineer, engineer for metrology, engineer in the organization of operation and repair, engineer of production preparation;
- Electrical technicians: electromechanician, ship electromechanician, electromechanician of the vessel electrical equipment, electromechanician of the underwater vehicle, group transloading machines electromechanician, electromechanician of lifting installations, electromechanics of underground sections, electromechanician-mentors, telecommunication electromechanincian, electromechanician dispatcher, district electrician, shopfloor electrician, electricsman;
- technical specialists in the field of extractive industry and metallurgy, technician-electromechanicians mining;

- ships specialists: electromechanician of the group fleet, electromechanics of the linear fleet, mechanic (electromechanician) (ship) – skipper, authorized to accept ships from shipbuilding factories.

The basic competencies determined by the educational qualification characteristics of electromechanics bachelor include the following: social-personal, general-scientific, instrumental, general-professional and specialized-professional.

National Center for Educational Statistics of the US Department of Education branch of knowledge 0507 – Electrical engineering and electromechanics are divided into separate branches of knowledge: 15.03 – Electrical Engineering Technologies and 15.04 – Electromechanical Instrumentation and Maintenance Technologies [5]. The following areas of training are included in the field of knowledge 15.04: biomedical technologies, electromechanical technologies and electromechanical engineering, measuring instruments, robotics technologies, automation technologies, electromechanical measuring instruments and their servicing.

Thus, in the training of electromechanical engineers, four branches of engineering are involved: mechanical and electrical engineering (with a common core of electromechanics), electronic engineering and automation. The common use of the theory, methods and means of these industries leads to the emergence of a combined field of engineering – mechatronics (Fig. 2).

Uday Shanker Dixit defines mechatronics as "a synergetic integration of mechanical engineering with electrotechnics and / or electronics, and possibly with other disciplines, for the purpose of designing, manufacturing, operating and maintaining a product" [6, p. 75].

Despite the lack of a holistic study of the process of training engineers-electromechanics in domestic and foreign works, some components of this process were considered in a number of theses devoted to the training of electricians.

Giuzel S. Sagdeeva distinguishes the general intellectual qualities of the engineer's personality on the operation of electrical devices: ability to concentrate attention, ability to allocate essential features, ability to make a deliberate decision in a difficult technical situation, ability to manage and organize the work of personnel, ability to work with schemes and drawings, content in the memory of devices, models and devices, the ability to self-improvement [39, p. 9].

These qualities of an engineer's personality are the result of the formation of intellectual competence, the acquisition of which provides the basis for: the development of students of all components of the content of education; solving various life and professional problems; overcoming stereotypes and patterns of thinking; development of abilities to flexible variational perception and assessment of events occurring; reflection and consolidation of the experience of effective activity and success in a competitive environment. Sagdeeva's intellectual competence is defined as "metastability, which, by defining the degree of development by the subject of a certain domain, is characterized by a special type of organization of subject-specific knowledge and effective decision-making strategies in this subject area", distinguishing in its structure the following components: motivational, cognitive and metacognitive. The components of the motivational component are: readiness of students for self-education and development;

the presence of motives that lead to cognitive activity; personality orientation. The cognitive component includes the ability to work with information: the ability to search, structure, transform, transfer information from one method of encoding to another; ability to make generalizations, conclusions, to highlight the main thing; the ability to compile cognitive schemes of mental activity, algorithms for solving problems. The metacognitive component is represented by the skills and abilities of intellectual self-management and self-organization: it is the ability to set goals, to plan, evaluate, control the cognitive activity, the ability to self-assess and reflexive analysis.

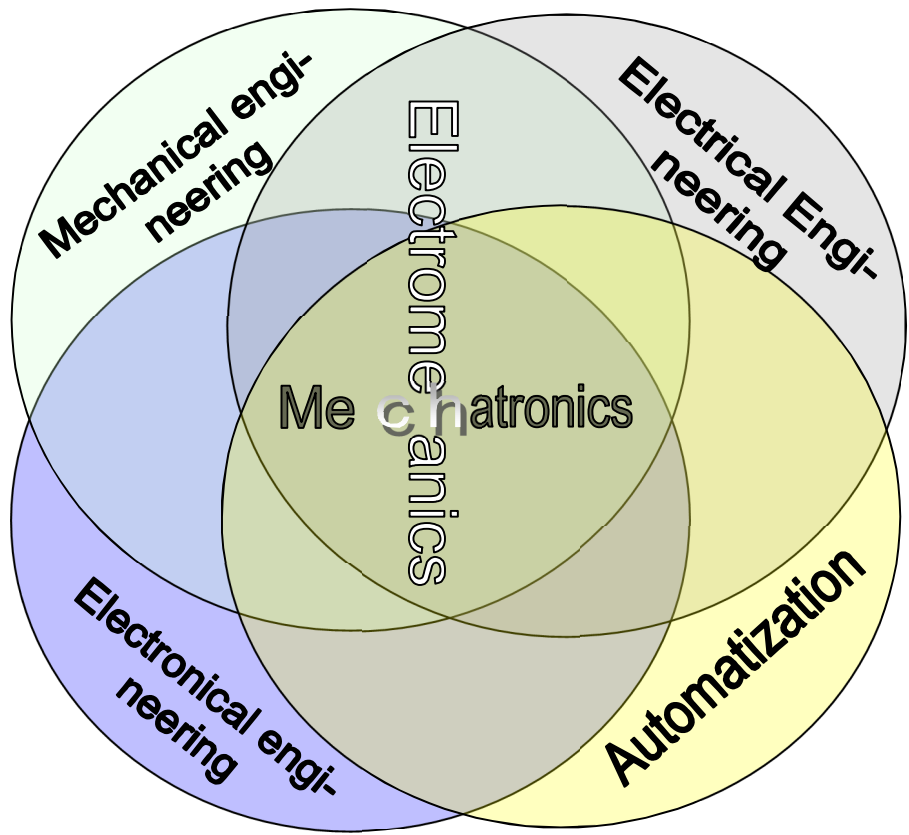

**Fig. 2.** Mechatronics as a combined branch of engineering

The conditions of intellectual competence of future electricians' development are:

1. simulation of intellectual and developmental situations in accordance with the psychological patterns and mechanisms of development of intellectual competence, taking into account the features of the future profession;
2. inclusion of students in various types of research activities aimed at the development and enrichment of invariant intellectual structures of the individual; improvement of student research methods based on the disclosure and formation of individual styles of intellectual activity;
3. development of psychological and pedagogical support of the process of training future electricians, which implements stimulating, diagnostic and corrective functions [39, p. 12-16].

The development of intellectual competence contributes to the formation of professional electrical thinking directed, according to Larisa N. Vishniakova, to the knowledge, understanding and transformation of electrotechnical objects, phenomena, processes and relations: "the essence of professional electrical engineering is manifested in its laws, namely, in natural conformance (based on the experience of human interaction with the biosphere, technosphere, society), cultural correspondence (associated with the mastery of general-professional and special knowledge and skills that are presented to the profession of social order of society) and the optimum combination of (relatively stable asymmetric harmony or complementarity) natural intuition of foresight and intellectual discipline in the performance of cognitive training and professional action" [44].

In its development, the professional electrical engineering of the student passes the following levels: elementary-empirical (zero), student, methodical, search. The transition of professional electrical thinking from one level to another is associated with transitions in intellectual development: electrical engineering – electrotechnical education – professional competence – electrical engineering and technological culture.

Elena V. Shishchenko [40] and Aleksandr V. Gamov [13] considered the formation and development of professional competencies of students on the basis of interdisciplinary integration. According to Shishchenko, "the interdisciplinary integration of knowledge contributes to competent education, person-oriented technologies of learning, technology of developmental learning, project method, block-module training, contextual training, wide-profile training of specialists, adult learning technology , oriented to the perception and assimilation of knowledge, representing a coherent system; on the formation of skills to perform certain operations, tasks (including research, creative), associated with their professional activities" [40, p. 5]. Integration of electrical engineering disciplines (theoretical electrical engineering, electrical measurements, electronic equipment, electric machines, electric drive and converters) contributes to solving the contradiction between the fast-changing elemental base of electrical installations and aggregates, which are constantly complicated by their algorithmic structure and circuitry, on the one hand, and some conservatism of typical programs and tutorials that contain information on individual, often outdated, electrical installations, on the other hand [40, p. 7].

Dixit takes notice that modern training engineers and electricians must be based on a top-down approach in which first provided a general idea of the final product, though not in great detail the form and then studied in detail subsystem system. This is due to the fact that such training involves many disciplines from different fields of engineering, so students should get an idea of how they will be integrated, "the integration of different disciplines is an essential part mechatronics" [6, p. 86].

Gamov adds that "the integrative approach reveals the possibilities of developing professional competences on the basis of integration: general-professional, special disciplines and information technologies; technologies of problem and modular learning; methods of classical calculation and modeling of electrodynamic systems" [13, p. 11].

Thus, the level of the formation of professional competence of masters of electrical engineering direction Galina Iu. Dmukh [7] determines the degree of development of the following competencies: research (the collection, analysis, processing and systematization of scientific and technical information, the ability to participate in all phases of research, the ability to use the achievements of science and technology, advanced national and foreign experience); operational (ability to carry out examination of technical documentation, supervision and control over the state of technological processes and operation of equipment, ability to effectively use natural resources, materials and energy); design (the ability to carry out a comprehensive technical and economic analysis, knowledge of methods for conducting technical calculations and determination of the economic efficiency of research and development, knowledge of the principles of work, technical, design features of the developed and used technical means); production-technological (knowledge of technology for the design, production and operation of products and facilities for technological equipment); organizational and managerial (interaction with specialists of the related profile). From the experience of masters of electromechanics at the Royal Institute of Technology (Sweden), Mats Hanson came to the conclusion that the most useful project in the teaching of mechatronics is the design-oriented approach [14].

The separation of the competences of the future specialist in the electromechanical profile in the process of simulation of professional training, according to Natalia P. Motorina [31], should be carried out on the basis of a specialist's model, the components of which are:

— identification of a range of main tasks solved by a modern electromechanician (model of activity);
— definition of the complex necessary for a specialist knowledge, skills and professional skills based on the model of activity (model of training);
— clarification of the necessary professional qualities of the specialist (model of personal qualities);
— preparation for the acquisition of perspective directions of development for this specialty, based on the forecast of its development for the next 15-20 years (model of the prospects of the specialty).

According to the results of modeling, the design and implementation of the profile education system (Sergei N. Kashkin [17] vocational training and retraining of specialists on the basis of the theory of continuous multi-level vocational education is carried out. Sergei A. Pchela [34] established the following pedagogical regularities of continuity, characteristic for the continuous training of specialists: structural, procedural and content continuity determine the content of educational programs, the content and quality of teaching and methodological provision of training, the level and quality of material and technical provision of training, the order and sequence of theoretical and practical training, the choice of forms and methods of teaching, types of educational activities and methods for diagnosing the level of professional training of specialists, the level of per training, training of teachers for the implementation of quality education programs. Elena A. Dragunova [8] notes that in this approach, the quality of training can be improved, in particular, through the use of modern software for distance learning and the possibilities of Internet technologies.

The purpose of continuous multi-level vocational education is the training of skilled professionals capable of navigating in ever-changing reality, mastering new modern technologies, implementing them in practice and successfully mastering fundamentally new areas and activities. Successfully self-realizing and feeling comfortable in a modern society, as well as ensuring its sustainable development will be able professionals who can mobilize themselves to improve themselves and transform their professional reality in accordance with the requirements of time and modern society. Tatiana B. Kotmakova [18] defines one of the main professional characteristics of the future specialist, which increases his competitiveness in the labor market - personal mobility - as an integrative quality of the future specialist, which manifests itself in the formed motivation to study, the ability to work in an effective way communication and allows you to stay in the process of active creative self-development.

Increasing competitiveness requires mastering by the future specialist a set of knowledge, skills necessary to active creative professional development, continuous self-improvement and training during the work activity. Therefore, an important task for the professional training of future engineers-electromechanics is not so much the acquisition of ready-made knowledge, as mastering the methods of independent cognitive activity. Maiia H. Hordiienko [15] emphasizes that under accelerated accumulation and obsolescence professionally significant information mastering abilities and skills of independent work enables future professionals to be constantly informed of the latest technologies in his professional field, equips achievements of world science and practice: "At the same time, professionally competent electromechanicians must solve the urgent national problem of energy conservation through the use of various technologies driven which provide the necessary modes of operation of electromechanical complexes. These technologies are implemented by a variety of converters, soft starters, microprocessor management, etc., a significant number of which are produced by foreign companies. To explore and use the best international experience on the latest developments, future electromechanical engineer must be able to independently find the information you need to read it in a foreign language is to possess abilities and skills of independent work with foreign professional literature" [15, p. 3].

Under these conditions, the problem of forming skills and abilities of independent work for future engineers becomes of particular importance in order to ensure their adaptation, self-realization and self-education in the modern conditions of the information society and integration into the world community. The purposeful formation of skills and abilities of independent work of bachelors of electromechanics should begin with fundamental training, which is based on mathematics, physics and informatics.

Tetiana V. Krylova indicates that mathematics as a basis for the study of fundamental, general technical and special disciplines provides wide opportunities for the development of logical thinking, algorithmic culture, the formation of skills to establish causal relationships, to substantiate statements, to model, etc.: "if the methodical system of education Mathematics of bachelors of electromechanics will take into account: the professional orientation of teaching mathematics; learning the beginnings of mathematical modeling in studying the general course of higher mathematics and special mathematical courses; solving problems of special content at the final stage of studying the disciplines of the mathematical cycle; methods, methods and means of activating the independent educational and cognitive activity of students in the study of mathematics; application of means of new information technology training in solving applied problems in the process of studying the general course of mathematics and special mathematical courses; level differentiation and individualization of teaching mathematics students of technical specialties; organization of independent work of students and control over its implementation, this will ensure the implementation of modern requirements for the mathematical preparation of students, promote their mental development, preparation for self-education in conditions of continuing education" [19].

Aleksandra N. Lavrenyna [20] proposes to fill a physics course by taking into account the profile of the training of future specialists, in particular, by analyzing the connections of the electrodynamics of the course in physics with the general technical discipline "Theoretical Foundations of Electrical Engineering" and the special discipline "Electric Machines" with the purpose of determining the role and places of physical knowledge in the system of vocational education of students of electrotechnical specialties.

Svetlana N. Potemkina [36] defined the general requirements for the professional training of an electrical engineer profile in the field of physics:

— to know and to be able to use the basic concepts, laws and models of mechanics, electricity and magnetism, oscillations and waves, quantum physics, statistical physics and thermodynamics;
— to know and to be able to competently solve complex tasks, which include tasks by type of activity;
— to know and to be able to use the methods of theoretical and experimental research in physics;
— to be able to evaluate the numerical order of quantities characteristic of different sections of science;
— to know and to be able to apply standard rules for constructing and reading drawings and diagrams;
— to know the principles of symmetry and conservation laws;
— to know about physical modeling.

Interdisciplinary and modeling skills are used in all components of the fundamental and professional training of the bachelor of electromechanics. A striking example of the use of interdisciplinary modeling is the methodology for the formation of environmental knowledge of future engineers-electromechanics in the process of teaching special disciplines, the author of which developed Iryna O. Soloshych, points out that "the involvement of students in the solution of problem-oriented nature of simulated produc-tion situations using interactive and informational methods promotes the effective development of their professional interests, motivation to master the future specialty" [42, p. 12].

Roman M. Sobko offers the following principles for the integrative use of ICT facilities in the training of students of electrical and electromechanical specialties, the main of which are the principles:

— the purposeful use of ICT tools in the professional training of specialists, which provides methodological, psychological, pedagogical and methodological substantiation of the content of ICT education;
— professional orientation of ICT training;
— continuity of use of ICT at all stages of vocational training;
— the degree and systematic formation of the ITC competence of a future specialist
— awareness of the use of ICTs in solving professional problems;
— modeling of phenomena and processes of professional activity using ICT tools [34, p. 9-10].

The implementation of the latter two principles is possible provided that the future specialists prepare for the engineering experiment, which Raisa E. Mazhirina [21] defines as the property of the individual to manage the active cognitive process associated with the analysis of qualitative and quantitative characteristics of industrial objects. The training of future engineers for independent studies, including the development of techniques and techniques of experiment, is an essential part of the professional training of an engineer, whose production activity is associated with constant analysis and directed change of technical and natural systems. Considering that training in electromechanicians takes up a significant place in the field of quick-change engineering – electronic, – the use of ICT for modeling phenomena and processes of professional activity is necessary both in the process of professional training and in the process of professional activity, which necessitates the use of mobile modeling tools.

In the teaching of electrical engineering disciplines using ICT, Natalia P. Fiks [11] suggests using automated teaching and learning complexes, which include computer-based learning tools: textbooks, training generators, virtual laboratories, diagnostic tools and automated systems modeling. An example of such a complex is developed by Natalia G. Pankova [33] a complex of software and information support for the process of teaching electrical engineering disciplines, consisting of training manuals on the simulation and calculation of electrical circuits, methodological instructions for a laboratory workshop using ICT, programs, guidelines and control tasks for calculation and graphic works, test control system of success, system of training classes on the basis of ICT. The highest level of automation of the teaching-methodical complex is realized by Maksim A. Polskii [35] a combined didactic interactive program system that provides the organization of reproductive (recognition and reproduction) and productive heuristic educational and cognitive activity of students in the conditions of gradualness and completeness of studying with a closed directional automatic control. Among the conditions for the effectiveness of the organization of the educational process using such complexes, the researcher calls the high level of ICT competencies of teachers and

students – in particular, the ability to work with universal software systems for modeling.

Considering the educational perspectives of applied mechatronics in the context of the integration of traditional topics of mechanical, electrical and computer engineering, C. J. Fraser et al. [12] offer the following sections of the curriculum: system engineering; microprocessor technology; digital electronics; digital and analog interfaces; digital communications; software development; Subordinate management of electric, pneumatic and hydraulic systems; the theory of automatic control.

Joshua Vaughan, Joel Fortgang, William Singhose, Jeffrey Donnell, and Thomas Kurfess [43] offer an integrative course "Creative Solutions and Design" aimed at the formation and development of students of mechatronic and communicative competences. The authors, emphasizing the importance of working in the team, note that team work can not equally develop students' competencies in all relevant fields, so they share work in accordance with their own comfort and abilities. In order to avoid this at the beginning of the course, it is expedient for each student to give an individual project, and in the second half of the course students are involved in team projects.

Yu Wang, Ying Yu, Chun Xie, Huiying Wang, and Xiao Feng [45] described 4 units of practical training at the CDHAW Center at Tongji University (China):

1. pre-training block includes study of the basics of mechanical, electrical and electronic engineering;
2. the block of fundamental training involves laboratory work, in which students check the laws of mechanics, physics, materials science, electrical engineering, etc.;
3. a block of specialized training involves laboratory work using controls, sensors, drives, controllers, microprocessors, etc.;
4. the unit of advanced training involves the student's independent work on projects.

The basic requirements for the professional training of specialists in electromechanics, formulated by the survey of employers, leads Maurice W. Roney [37, p. 26]:

1. Preparation should be fundamental: the emphasis should be more on the general principles of the work of electromechanical systems than on the application of these principles.
2. Communicative skills are extremely important in the work of electronics technicians, so they should be given special attention in the training program.
3. Study of the interconnection of electrical and mechanical elements of systems and devices should occupy a central place in specialized technical courses. Wherever possible, electrical and mechanical principles should be studied together, not alone.
4. Principles of electrical and mechanical physics are the main tools in the work of electronics technicians and any technical training should develop the skills of analytical thinking for which these tools are fundamental. In addition, there is an increasing need for techniques for working with new branches of application of other physical sciences such as: optical equipment, thermal power plants, hydraulic and pneumatic controls, as well as a wide range of measuring instruments.

To implement these requirements is proposed [37, p. 10]:

1. The main subjects that should be given the greatest attention are:

   - physics – of the applied type (should not be classical physics);
   - mathematics – through applied calculus;
   - communications – drafting, sketching, composition, report writing;
   - industrial electronics – regardless of the area in which the technician might be working, a good working knowledge of electronic devices, circuits, instruments and system is required.

2. The training program should also include material from the sections:

   - light and optics;
   - high vacuum techniques;
   - engineering materials and stress analysis;
   - chemistry, particularly from the viewpoint of corrosion;
   - economics – as applied to industrial situations in design and application;
   - mechanism and basics of mechanical design;
   - transducers for various types of instrumentation;
   - controllers and industrial control;
   - fundamentals of computers.

3. It is very important to have an exact observation: by carefully observing, the technician must be able to analyze and synthesize. Although these two abilities may not develop intentionally in a particular course, they should be developed in all laboratory and classroom activities. Competence in these areas can be more important than just technical abilities.
4. The skills of manual labor with basic tools are also important.
5. If practicable, the training program should be no more than two years old.

In the training of electronics technicians, Roney proposes to follow a model that has a four-component structure (Fig. 3). In the center of the model – a student, on the development of the personality which must be sent all the efforts of pedagogues. To the teaching staff, Roney proposes a requirement for competence in more than one discipline in order to provide interdisciplinary connections and integration of academic disciplines [38, p. 20].

The development of the communicative competence of a future specialist should be supported by all pedagogues: the pedagogue "should not reduce his teaching function to writing mechanics. Instead, he must be able to distinguish the specific needs of students at each stage of the program. He must understand that without special communicative skills, the technician will be poorly trained to perform production functions" [38, p. 22].

But the most important requirement for pedagogues preparing future specialists in electromechanics, Roney considers "his production experience, which should be significant and as modern as possible. One of the main problems of teaching is the lagging content of training from the current state of development of production" [38, p. 22]. The prestige of the educational institution, according to the author, largely depends on the extent to which the qualifications of the pedagogues correspond to the current state of development of production.

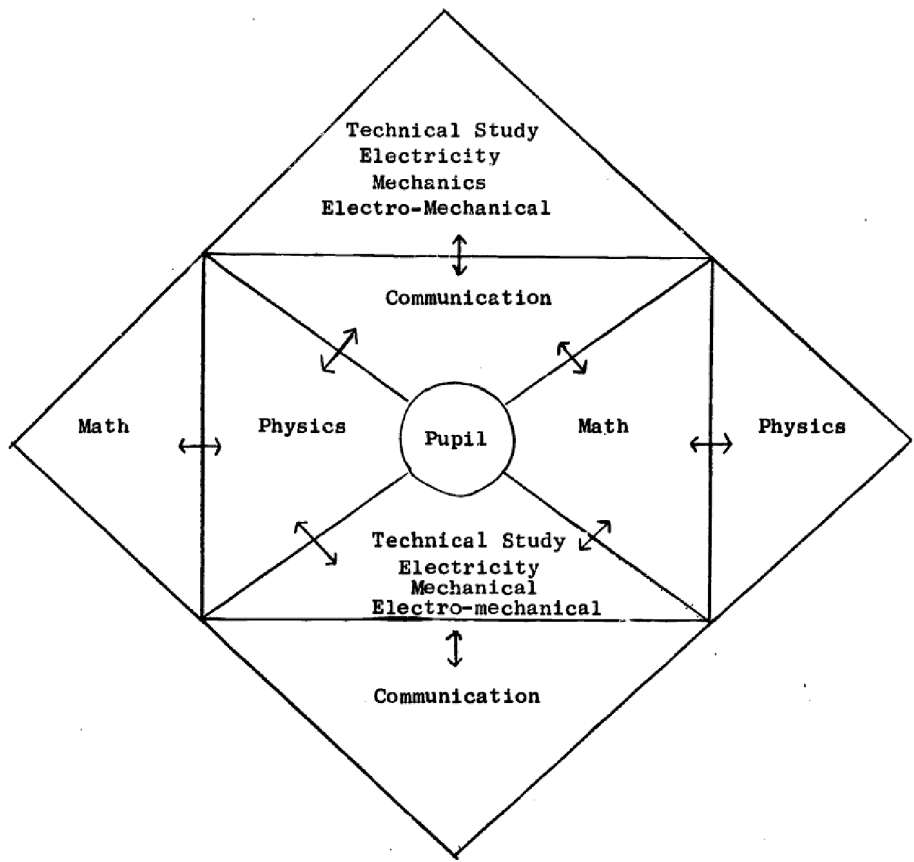

**Fig. 3.** Technicians-electromechanics training model (by [38, p. 21])

## 3 General and professional competence of electrical engineers in United States and Canada

The professional training of electromechanical engineers in the United States (The Bachelor of Science in Electro-Mechanical Engineering Technology – BSEMET), according to the ABET (Accreditation Board for Engineering and Technology), has been providing since the early 1990s. according to the related branch of science (Electromechanical Engineering Technology, Engineering Technology: Electro-Mechanical Concentration, Electromechanical Engineering Technology Concentration in Engineering Technology) in accordance with the developed ABET accreditation criteria for training programs for engineers, which identified the necessary requirements for the program of electromechanical training (Electromechanical Engineering Technology).

The requirements of ABET [4, p. 15] clearly distinguish professional activities of electronics and electromechanics. The production functions of the technique-electromechanics include the construction, installation, use and operation and / or maintenance of electromechanical equipment and software. The bachelor of electromechanics will, design, development and management of electromechanical systems.

Technician-electromechanics should have the following competencies:

1. use computer-aided drafting or design tools to prepare graphical representations of electromechanical systems;
2. use circuit analysis, analog and digital electronics, basic instrumentation, and computers to aid in the characterization, analysis, and troubleshooting of electromechanical systems;
3. use statics, dynamics (or applied mechanics), strength of materials, engineering materials, engineering standards, and manufacturing processes to aid in the characterization, analysis, and troubleshooting of electromechanical systems;

Graduates of baccalaureate degree programs must also demonstrate competency to:

4. use appropriate computer programming languages for operating electromechanical systems;
5. use electrical / electronic devices such as amplifiers, motors, relays, power systems, and computer and instrumentation systems for applied design, operation, or troubleshooting electromechanical systems;
6. use advanced topics in engineering mechanics, engineering materials, and fluid mechanics for applied design, operation, or troubleshooting of electromechanical systems;
7. use basic knowledge of control systems for the applied design, operation. or troubleshooting of electromechanical system;
8. use differential and integral calculus, as a minimum, to characterize the static and dynamic performance of electromechanical systems;
9. use appropriate management techniques in the investigation, analysis, and design of electromechanical systems.

There are eight ABET Accreditation Criteria common to all Engineering Training Areas [4, p. 1-5].

The first criterion defines the requirements for the process and the results of the professional training of students; Separately, it is indicated the need to monitor the training of each student in order to facilitate the achievement of educational goals.

The second criterion defines the requirements for the educational objectives of the training program.

The third criterion defines two groups of competencies that students must acquire in order to achieve the objectives of the training program. The first group defines broad-sighted activities related to: the use of different resources; Innovative use of new processes, materials or technologies; execution of standard operating procedures. The second group defines, in the narrow sense, activities that involve limited resources, new ways of using traditional processes and materials, and the implementation of basic operating procedures.

For bachelors of engineering, there are such competencies:

1. an ability to select and apply the knowledge, techniques, skills, and modern tools of the discipline to broadly-defined engineering technology activities;
2. an ability to select and apply a knowledge of mathematics, science, engineering, and technology to engineering technology problems that require the application of principles and applied procedures or methodologies;
3. an ability to conduct standard tests and measurements; to conduct, analyze, and interpret experiments; and to apply experimental results to improve processes;
4. an ability to design systems, components, or processes for broadly-defined engineering technology problems appropriate to program educational objectives;
5. an ability to function effectively as a member or leader on a technical team;
6. an ability to identify, analyze, and solve broadly-defined engineering technology problems;
7. an ability to apply written, oral, and graphical communication in both technical and nontechnical environments; and an ability to identify and use appropriate technical literature;
8. an understanding of the need for and an ability to engage in self-directed continuing professional development;
9. an understanding of and a commitment to address professional and ethical responsibilities including a respect for diversity;
10. a knowledge of the impact of engineering technology solutions in a societal and global context;
11. a commitment to quality, timeliness, and continuous improvement.

The requirement of continuous improvement of the training program is the basis of the fourth criterion. It is proposed to apply appropriate methods for assessing and analyzing student achievements. The obtained results should be used systematically as inputs to continuously improve the training program.

The fifth criterion defines the general requirements for the curriculum:

— The mathematics program must develop the ability of students to apply mathematics to the solution of technical problems. Programs will include the application of integral and differential calculus or other mathematics appropriate to the student outcomes and program educational objectives;
— The technical content of the program must focus on the applied aspects of science and engineering and must represent at least 1/3 of the total credit hours for the program but no more than 2/3 of the total credit hours for the program. Include a technical core that prepares students for the increasingly complex technical specialties they will experience later in the curriculum. Develop student competency in the use of equipment and tools common to the discipline;
— The basic physical and natural science content of the program must include physical or natural science with laboratory experiences as appropriate to the discipline;

- Baccalaureate degree programs must provide a capstone or integrating experience that develops student competencies in applying both technical and non-technical skills in solving problems;
- When used to satisfy prescribed elements of these criteria, credits based upon cooperative / internships or similar experiences must include an appropriate academic component evaluated by the program faculty;
- An advisory committee with representation from organizations being served by the program graduates must be utilized to periodically review the program's curriculum and advise the program on the establishment, review, and revision of its program educational objectives. The advisory committee must provide advisement on current and future aspects of the technical fields for which the graduates are being prepare.

The sixth criterion defines the requirements for teachers, the main is the availability of experience and level of education, corresponding to the expected input of the teacher in the training program. Teacher competence is assessed by education, professional qualification and certification, professional experience, current professional development, discipline, teaching efficiency and communication skills. Together, all teachers should cover all components of the training program.

The staff involved in the training program should be in sufficient quantity to maintain continuity, stability, control, student interaction and counseling. The staff should have sufficient responsibility and authority to improve the curriculum by identifying and reviewing educational goals and learning achievements, as well as for implementing a training program that will help improve student achievement.

The seventh criterion defines the requirements for the means of support (facilitation) of the learning process:

- classrooms, offices, laboratories, and associated equipment must be adequate to support attainment of the student outcomes and to provide an atmosphere conducive to learning;
- modern tools, equipment, computing resources, and laboratories appropriate to the program must be available, accessible, and systematically maintained and upgraded to enable students to attain the student outcomes and to support program needs;
- students must be provided appropriate guidance regarding the use of the tools, equipment, computing resources, and laboratories available to the program;
- the library services and the computing and information infrastructure must be adequate to support the scholarly and professional activities of the students and faculty.

The eighth criterion defines the level of support for a training program from a parent institution and management that is sufficient to ensure the quality and integrity of the training program:

- resources including institutional services, financial support, and staff (both administrative and technical) provided to the program must be adequate to meet program needs;
- the resources available to the program must be sufficient to attract, retain, and provide for the continued professional development of a qualified faculty;

— The resources available to the program must be sufficient to acquire, maintain, and operate infrastructures, facilities and equipment appropriate for the program, and to provide an environment in which student outcomes can be attained.

Standards for the training of electromechanical engineers, proposed by the Department of Education, Ontario Colleges and Universities (Canada) [32], contain three components: Vocational standard – analogue of special professional competencies of the domestic standard, Generic employability skills standard – an analogue of general-professional, instrumental (partly) and general-knowledge (partly) competencies of the domestic standard, and General education standard – an analogue of socio-personal, instrumental (partly) and general (partly) competencies of the domestic standard.

As a result of mastering the Vocational Standard, the following competencies should be formed for graduates:

— fabricate mechanical components and assemblies, and assemble electrical components and electronic assemblies by applying workshop skills and knowledge of basic shop practices in accordance with applicable codes and safety practices;
— analyse, interpret, and produce electrical, electronic, and mechanical drawings and other related documents and graphics necessary for electromechanical design;
— select and use a variety of troubleshooting techniques and test equipment to assess electromechanical circuits, equipment, processes, systems, and subsystems;
— modify, maintain, and repair electrical, electronic, and mechanical components, equipment, and systems to ensure that they function according to specifications;
— apply the principles of engineering, mathematics, and science to analyse and solve design and other complex technical problems and to complete work related to electromechanical engineering;
— design and analyse mechanical components, processes, and systems through the application of engineering principles and practices;
— apply principles of mechanics and fluid mechanics to the design and analysis of electromechanical systems;
— design, analyse, build, and troubleshoot logic and digital circuits, passive AC and DC circuits, and active circuits;
— design, select, apply, integrate, and troubleshoot a variety of industrial motor controls and data acquisition devices and systems;
— design, analyse, and troubleshoot microprocessor-based systems;
— install and troubleshoot computer hardware and high-level programming to support the electromechanical engineering environment;
— analyse, program, install, integrate, and troubleshoot automated systems including robotic systems;
— establish and maintain inventory, records, and documentation systems;
— assist in project management by applying business principles to the electromechanical engineering environment;
— select for purchase electromechanical equipment, components, and systems that fulfill the job requirements and functional specifications;
— specify, coordinate, and conduct quality-control and quality-assurance programs and procedures;

- perform all work in accordance with relevant law, policies, codes, regulations, safety procedures, and standard shop practices;
- develop personal and professional strategies and plans to improve job performance and work relationships with clients, coworkers, and supervisors [32, p. 6–7].

*Generic employability skills* standard defines the following competencies:

- communicate clearly, concisely, and correctly in the written, spoken, and visual form that fulfills the purpose and meets the needs of the audiences;
- reframe information, ideas, and concepts using the narrative, visual, numerical, and symbolic representations which demonstrate understanding;
- apply a wide variety of mathematical techniques with the degree of accuracy required to solve problems and make decisions;
- use a variety of computer hardware and software and other technological tools appropriate and necessary to the performance of tasks;
- interact with others in groups or teams in ways that contribute to effective working relationships and the achievement of goals;
- evaluate her or his own thinking throughout the steps and processes used in problem solving and decision making;
- collect, analyze, and organize relevant and necessary information from a variety of sources;
- evaluate the validity of arguments based on qualitative and quantitative information in order to accept or challenge the findings of others;
- create innovative strategies and / or products that meet identified needs;
- manage the use of time and other resources to attain personal and / or project related goals;
- take responsibility for her or his own actions and decisions;
- adapt to new situations and demands by applying and / or updating her or his knowledge and skills;
- represent her or his skills, knowledge, and experience realistically for personal and employment purposes [32, p. 27].

Goals and Broad Objectives of *General Education*:

- Aesthetic Appreciation: understand beauty, form, taste, and the role of the arts in society;
- Civic Life: understand the meaning of freedoms, rights, and participation in community and public life;
- Cultural Understanding: understand the cultural, social, ethnic, and linguistic diversity of Canada and the world;
- Personal Development: gain greater self-awareness, intellectual growth, well-being, and understanding of others;
- Social Understanding: understand relationships among individuals and society;
- Understanding Science: appreciate the contribution of science to the development of civilization, human understanding, and potential;

— Understanding Technology: understand the interrelationship between the development and use of technology and society and the ecosystem;
— Work and the Economy: understand the meaning, history, and organization of work; and of working life challenges to the individual and society [32, p. 46-48].

Specialists of Human Resource Systems Group [16] determine two groups of competencies that can be formed at one of five levels (1 – basic, 5 – expert):

1. *General Competencies* includes:

   - writing skills (at level 4 – writes on complex and highly specialized issues);
   - analytical thinking (at level 4 – applies broad analysis);
   - interactive communication (at level 4 – communicates complex messages);
   - problem solving (at level 4 – solves complex problems);
   - planning and organizing (at level 4 – plans and organizes multiple, complex activities);
   - team leadership (at level 3 – builds strong teams);
   - critical judgment (at level 4 – formulates broad strategies on multi-dimensional strategic issues);
   - visioning and alignment (at level 3 – aligns program / operational support);

2. *Technical Competencies* includes:

   - calibration / mathematics (at level 4 – calculates using multiple steps and operations);
   - working with tools and technology (at level 4 – welds, repairs, and fabricates equipment or machinery);
   - building & construction design (at level 4 – demonstrates advanced knowledge and ability, and can apply the competency in new or complex situations; guides other professionals);
   - electrical systems maintenance and repair (at level 4 – demonstrates advanced knowledge and ability, and can apply the competency in new or complex situations; guides other professionals);
   - electrical / electronics engineering (at level 5 – demonstrates expert knowledge and ability, and can apply the competency in the most complex situations; develops new approaches, methods or policies in the area; is recognized as an expert, internally and / or externally);
   - electrical equipment operation (at level 5 – expert).

Competence matrix for the sector electronics / electrical engineering [1, p. 14-15], developed within the framework of the European project VQTS II (Vocational Qualification Transfer System), covers 8 groups of competencies, each of which is defined at 3 or 4 levels:

1. planning, mounting and installing electrical and electronic systems;
2. inspecting and configuring electrical and electronical systems and machines in industrial appliances;

3. installing and adjusting electrical components and electronic systems;
4. designing, constructing and modifying electrical / electronic wirings / circuit boards, control circuitries and machines including their interfaces;
5. developing custom designed electrical / electronic systems;
6. supervising and supporting work and business processes;
7. installing, configuring modifying and testing of application software for the programming of electrical / electronic installations;
8. diagnosing and repairing of electrical / electronic systems and equipment.

## 4       Conclusions

Summarizing the experience of electrical engineers professional training in Ukraine and abroad makes it possible to determine the main directions of their professional training modernization:

1. transition to competence-oriented training standards;
2. development of integrated training programs for "technician-electromechanic engineer-electromechanic" on the basis of the National Qualifications Framework;
3. development of professional standards of training specialists in the field of mechatronics for the metallurgical and mining industry;
4. ensuring continuous training and retraining of electrical engineers based on the use of modern ICT tools.